# Superconductivity Induced by Breaking Te$_2$ Dimers of AuTe$_2$

Kazutaka Kudo[1,*], Hiroyuki Ishii[1], Masaya Takasuga[1], Keita Iba[1], Seiya Nakano[1], Jungeun Kim[2], Akihiko Fujiwara[2], and Minoru Nohara[1]

[1]*Department of Physics, Okayama University, Okayama 700-8530, Japan*
[2] *SPring-8/JASRI, Sayo, Hyogo 679-5198, Japan*

Mineral calaverite AuTe$_2$ is a layered compound with an incommensurately modulated structure. The modulation is characterized by the formation of molecular-like Te$_2$ dimers. We have found that the breaking of Te$_2$ dimers that occurs in Au$_{1-x}$Pt$_x$Te$_2$ results in the emergence of superconductivity at $T_c$ = 4.0 K.

The breaking and making of chemical bonds are essential for designing materials. Hoffmann and Zheng developed a theory of bond manipulation in ThCr$_2$Si$_2$-type AT$_2$X$_2$ (A = alkali, alkali-earth, or rare-earth element, T = transition element, X = group 14 or 15 element).[1] AT$_2$X$_2$ is formed by stacking covalently bonded T$_2$X$_2$ layers composed of edge-sharing TX$_4$ tetrahedra, and ionic A atoms. Interlayer X-X covalent bonds can be tuned by filling the $d$ band in T elements.[1] If the Fermi level $E_F$ at the $d$ band lies below the antibonding ($\sigma^*$) orbitals of X-X, the molecular-like X-X dimer is formed, while the X-X dimer is broken if $E_F$ is located above the $\sigma^*$ orbital. Their theoretical idea has been experimentally realized by the chemical doping of several AT$_2$X$_2$ compounds, which include doped CaFe$_2$As$_2$,[2,3] SrCo$_2$Ge$_2$,[4] and LaCo$_2$Ge$_2$.[5] A drastic change in the electronic state occurs along with the making and breaking of molecular-like dimers. For example, making interlayer As-As dimers in CaFe$_2$(As$_{1-x}$P$_x$)$_2$[2] and Ca(Fe$_{1-x}$Rh$_x$)$_2$As$_2$[3] suppresses the antiferromagnetic fluctuation, resulting in the disappearance of superconductivity. Breaking interlayer Ge-Ge dimers in SrCo$_2$(Ge$_{1-x}$P$_x$)$_2$[4] and LaCo$_2$(Ge$_{1-x}$P$_x$)$_2$[5] induces the Stoner instability, leading to a ferromagnetic order. Moreover, a recent report has shown that intralayer Ge-Ge dimers are formed in BaNi$_2$(Ge$_{1-x}$P$_x$)$_2$, and breaking them causes a softening of phonons, inducing the emergence of superconductivity at 2.9 K.[6] The effect of tuning dimers has been suggested in pyrite: $Ir_{0.94-x}Rh_xSe_2$ exhibits the maximum superconducting transition temperature $T_c$ = 9.6 K when the intradimer distance of Se-Se becomes the longest.[7] Thus, the emergent phenomenon of novel electronic phases triggered by making or breaking dimers would be expected in a wide variety of compounds.

*E-mail: kudo@science.okayama-u.ac.jp

Calaverite $AuTe_2$ is a natural mineral that includes $Te_2$ dimers.[8] The average structure is a monoclinically distorted $CdI_2$-type structure with a space group $C2/m$ (No. 12) composed of layers of distorted $AuTe_6$ octahedra.[9] Each Au atom is surrounded by six Te atoms with two short (2.67 Å) and four long (2.98 Å) Au-Te bonds. In this average structure, the Te atoms form zigzag chains with an interatomic distance of 3.20 Å, as shown in Fig. 1(a). In the actual structure, however, the incommensurate modulation ($\mathbf{q}$ = –0.4076$\mathbf{a}^*$ + 0.4479$\mathbf{c}^*$) induces the breaking up of the zigzag chains and the formation of isolated $Te_2$ dimers with a short interatomic distance of 2.88 Å,[10] as shown in Fig. 1(b). Initially, the occurrence of the incommensurate modulation was interpreted in terms of the mixed-valence state of $Au^+$ ($5d^{10}$) and $Au^{3+}$ ($5d^8$).[10] Later, however, X-ray photoelectron spectroscopy (XPS) revealed the homogeneous monovalent $Au^+$ ($5d^{10}$) of $AuTe_2$.[11] Au $5d$ bands lie several electron volts below $E_F$, corresponding to fully occupied $5d$ bands.[11] This result has been supported by a first-principles calculation,[12] which shows that Au $5d$ bands are mainly located from -6.5 to -4.0 eV below $E_F$. The electron density of states (DOS) at $E_F$ is predominantly contributed by Te $5p$ bands, with a small contribution of the tail of Au $5d$ bands.[12] Thus, the $Te_2$ dimers appear to be a key factor in the structural and electronic properties of $AuTe_2$.

In this letter, we report that the superconductivity at $T_c$ = 4.0 K emerges upon the platinum doping of $AuTe_2$. Superconducting $Au_{1-x}Pt_xTe_2$ crystallizes in a trigonal (distortion-free) $CdI_2$-type structure, in which $Te_2$ dimers in $AuTe_2$ are broken into isolated Te atoms. This result suggests that dimer breaking is responsible for the emergence of superconductivity.

Polycrystalline samples of $Au_{1-x}Pt_xTe_2$ with nominal $0 \leq x \leq 0.40$ were synthesized by a solid-state reaction. First, stoichiometric amounts of the starting materials Au (99.99%), Pt (99.99%), and Te (99.99%) were mixed and pulverized. They were heated to 500ºC in an evacuated quartz tube and then cooled to room temperature. The product was then powdered, pressed into pellets, heated to 430–460ºC, and subsequently cooled to room temperature. The heating and cooling rates were both equal to 20ºC/h. The obtained samples were characterized at room temperature by powder X-ray diffraction (XRD) using a Rigaku RINT-TTR III X-ray diffractometer with Cu Kα radiation, and by synchrotron radiation powder XRD in the BL02B2 of SPring-8 [wavelength λ = 0.42004(1) Å]. Rietveld refinement was performed using the RIETAN-FP program.[13] The magnetization $M$ was measured using a Quantum Design MPMS. The electrical resistivity $\rho$ and the specific heat $C$ were measured using a Quantum Design PPMS.

The obtained samples of $x = 0.0$ were confirmed to be a single phase of calaverite $AuTe_2$ with a monoclinic average structure from the powder XRD data. However, those of $0.2 \leq x \leq 0.4$ could not be indexed in the same way as those of $x = 0.0$.[14] For instance, as shown in Fig. 2, the synchrotron powder XRD pattern of $x = 0.35$ could be indexed on the basis of the trigonal (distortion-free) $CdI_2$-type structure with the space group $P\bar{3}m1$ (No. 164).[15] The monoclinic and trigonal phases coexist at $x = 0.10$ and $0.15$.[14] The obtained crystallographic data of $Au_{0.65}Pt_{0.35}Te_2$ are summarized in Table I. It is found that the interatomic distance of Te atoms is uniform [3.2995(9) Å] and longer than that of a covalent $Te_2$ bond (2.91 Å in $FeTe_2$[10]). Thus, the $Te_2$ dimers in $AuTe_2$ are completely broken in $Au_{0.65}Pt_{0.35}Te_2$, as shown in Fig. 1(c).

We found that the breaking of the $Te_2$ dimer induces the emergence of superconductivity in $Au_{0.65}Pt_{0.35}Te_2$. As shown in Fig. 3(a), the electrical resistivity $\rho$ of $Au_{0.65}Pt_{0.35}Te_2$ shows a sharp drop below 4.2 K, characteristic of a superconducting transition. Zero resistivity was observed at 4.0 K, and the 10–90% transition width was estimated to be 0.09 K. $T_c$ decreases with increasing magnetic field, and disappears at 15 kOe. The magnetic field dependence of $\rho$, shown in Fig. 3(a), demonstrates that $Au_{0.65}Pt_{0.35}Te_2$ is a type-II superconductor. Figure 4 shows the temperature dependence of the upper critical field $H_{c2}$ determined from the midpoint of the resistive transition. $H_{c2}$ increases almost linearly with decreasing temperature down to 1.88 K and deviates from the curve expected from the Werthamer–Helfand–Hohenberg (WHH) theory[16]. We estimate a lower limit of the upper critical field at 0 K, $H_{c2}(0) = 12.9$ kOe, from the slope of $-dH_{c2}/dT = 4.6$ kOe/K at $T = T_c$. The Ginzburg–Landau coherence length $\xi_0$ was estimated to be 160 Å from $\xi_0 = [\Phi_0 / 2\pi H_{c2}(0)]^{1/2}$, where $\Phi_0$ is the magnetic flux quantum.

The bulk superconductivity in $Au_{0.65}Pt_{0.35}Te_2$ was evidenced by the temperature dependences of the magnetization $M$ and the specific heat $C$. The temperature-dependent $M$ shown in Fig. 3(b) exhibits a diamagnetic behavior below 4.0 K. The shielding and flux exclusion signals correspond to 71 and 31% of perfect diamagnetism, respectively. The electronic specific heat $C_e$ shown in Fig. 3(c) exhibits a clear jump at the superconducting transition. The assumption of an ideal jump at $T_c$ to satisfy the entropy conservation at the transition gives the estimates of $T_c$ = 3.84 K and $\Delta C/T_c$ = 8.60 mJ/molK$^2$.

Pristine AuTe$_2$ did not exhibit superconductivity. The ρ keeps decreasing with decreasing temperature down to 1.8 K and approaches a small residual resistivity of 2.9 μΩcm, as shown in the inset of Fig. 3(a). The metallic ρ(T) curve of AuTe$_2$ is consistent with the metallic character suggested by the XPS measurements[11] and the first-principles calculation.[12] In contrast, the ρ(T) curve of Au$_{0.65}$Pt$_{0.35}$Te$_2$ is almost flat with a large residual resistivity of 80 μΩcm, indicating that carrier density, as well as carrier scattering mechanisms, are largely altered by the Pt doping and resultant Te$_2$ dimer breaking.

The specific heat data indicate a strong enhancement in the DOS at $E_F$ for Au$_{0.65}$Pt$_{0.35}$Te$_2$. The standard analysis of the normal-state specific heat, whose results are shown in Fig. 5, yielded the electronic specific heat coefficient γ = 1.10(6) mJ/molK$^2$ and Debye temperature $\Theta_D$ = 182.2(4) K for the pristine AuTe$_2$, and γ = 5.48(7) mJ/molK$^2$ and $\Theta_D$ = 187.3(5) K for the Pt-doped Au$_{0.65}$Pt$_{0.35}$Te$_2$. It was found that Au$_{0.65}$Pt$_{0.35}$Te$_2$ acquires a fivefold enhancement in γ with a small change in $\Theta_D$ compared with AuTe$_2$. Using the obtained γ, we estimated $\Delta C/\gamma T_c$ to be 1.57 for Au$_{0.65}$Pt$_{0.35}$Te$_2$, which is larger than the value expected from the BCS weak-coupling limit ($\Delta C/\gamma T_c$ = 1.43). In contrast to the strong-coupling superconductivity that originates from soft phonons induced by chemical doping,[6,17] the change in $\Theta_D$ upon Pt doping is considerably small in the present system. The moderately strong-coupling superconductivity observed in Au$_{0.65}$Pt$_{0.35}$Te$_2$ is attributed to the enhanced electron-phonon coupling due to the enhanced DOS.[18] We failed to determine Pauli paramagnetic susceptibility because of the large core diamagnetism of Pt, Au, and Te, which results in a negative susceptibility in the normal state.

Our results show that the Pt doping of AuTe$_2$ induces the breaking of Te$_2$ dimers accompanied by a DOS enhancement. Here, two questions arise. One is why does dimer breaking occur upon Pt doping. The answer to this remains to be clarified. In the analogy to X-X dimers in AT$_2$X$_2$,[1] it is speculated that Te$_2$ dimers are broken by the filling up of σ$^*$ orbitals. The degree of band filling decreases with the Pt doping level of AuTe$_2$. At the same time, we expect that the band position increases with the Pt doping level, because the 5$d$ orbitals of Pt are considered to be higher than those of Au in energy.[1] When the latter dominates, $E_F$ increases with the Pt doping level, resulting in electron injection into σ$^*$ orbitals. Another question is what is the origin of the strong enhancement in DOS. In simple words, carrier doping resulting from the substitution of Pt for Au may enhance the DOS. Moreover, the breaking of Te$_2$ dimers may induce reconstruction in the band structure around

$E_F$. A similar result has been reported in Ca(Fe$_{1-x}$Rh$_x$)$_2$As$_2$, which exhibits a twofold increase in the Pauli paramagnetic susceptibility, and thus DOS at $E_F$, along with the breaking of As$_2$ dimers.[3] At this stage, however, we have not been able to ascertain whether this occurs. To address the above two questions, angle-resolved photoemission spectroscopy and a first-principles band calculation for Au$_{0.65}$Pt$_{0.35}$Te$_2$ are expected to be helpful.

In summary, the electrical resistivity, magnetization, and specific heat data indicate the emergence of superconductivity at $T_c$ = 4.0 K in Au$_{0.65}$Pt$_{0.35}$Te$_2$. The Rietveld refinement of the powder X-ray diffraction data confirms that Au$_{0.65}$Pt$_{0.35}$Te$_2$ crystallizes in the trigonal CdI$_2$-type structure, in which Te$_2$ dimers of AuTe$_2$ are completely broken. The breaking of Te$_2$ dimers is responsible for the emergence of superconductivity. Our results demonstrate that the approach of breaking dimers controlled by chemical doping is effective for exploring novel electronic phases.

**Acknowledgment**

Part of this work was performed at the Advanced Science Research Center, Okayama University. It was partially supported by a Grant-in-Aid for Scientific Research (C) (25400372) from the Japan Society for the Promotion of Science (JSPS) and the Funding Program for World-Leading Innovation R&D on Science and Technology (FIRST Program) from JSPS. The synchrotron radiation experiments performed at BL02B2 of SPring-8 were supported by the Japan Synchrotron Radiation Research Institute (JASRI; Proposal No. 2012B1037, 1055).

Table I. Rietveld-refined structural parameters of $Au_{0.65}Pt_{0.35}Te_2$. Space group $P\bar{3}m1$ (No. 164), $a = 4.1341(3)$ Å, $c = 5.1535(3)$ Å, $R_{wp} = 5.176\%$, and $R_p = 3.585\%$.

| Atom | site | g | x | y | z | $100U_{iso}$ (Å$^2$) |
|---|---|---|---|---|---|---|
| Au | 1a | 0.65 | 0 | 0 | 0 | 0.589(1) |
| Pt | 1a | 0.35 | 0 | 0 | 0 | 0.59(9) |
| Te | 2d | 1 | 1/3 | 2/3 | 0.7210(2) | 1.259(8) |

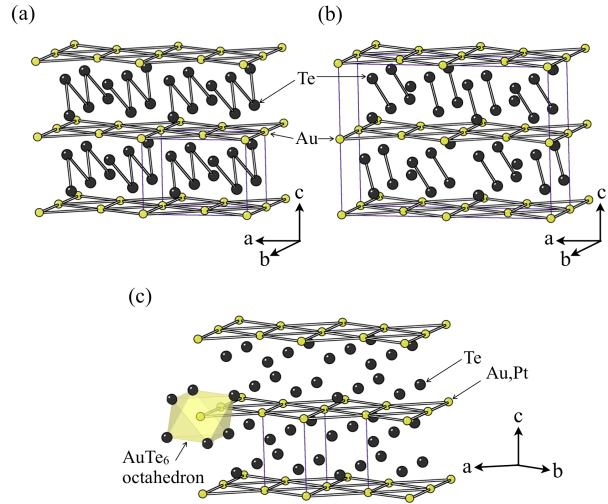

Fig. 1. (Color online) (a) Average structure of calaverite $AuTe_2$ with the space group $C2/m$.[9] Zigzag chains of Te atoms are formed. (b) Modulated structure of calaverite $AuTe_2$ approximated using a commensurate modulation vector ($\mathbf{q} = -1/2\mathbf{a}^* + 1/2\mathbf{c}^*$).[12] $Te_2$ dimers are formed. Note that the modulation of the real structure is incommensurate ($\mathbf{q} = -0.4076\mathbf{a}^* + 0.4479\mathbf{c}^*$).[10] (c) Structure of $Au_{0.65}Pt_{0.35}Te_2$ with the space group $P\bar{3}m1$ refined in Fig. 2. Te atoms are isolated. The solid lines indicate the unit cell.

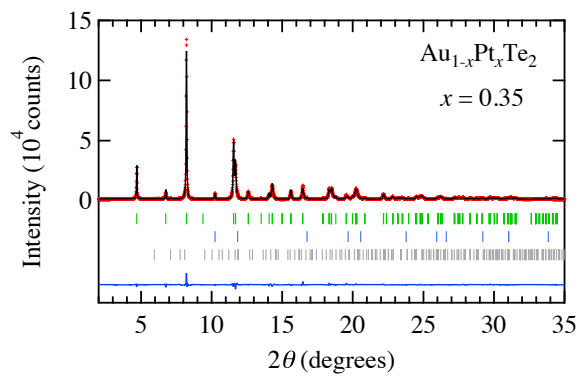

Fig. 2. (Color online) Synchrotron powder X-ray diffraction pattern measured at room temperature and its Rietveld refinement for $Au_{0.65}Pt_{0.35}Te_2$. The wavelength of X-ray is 0.42004(1) Å. The red circle, black line, and blue line indicate the observed, calculated, and difference profiles, respectively. The green, blue, and gray ticks indicate the Bragg diffraction positions calculated for $Au_{0.65}Pt_{0.35}Te_2$ with the space group $P\bar{3}m1$, for Au with $Fm\bar{3}m$, and for $TeO_2$ with $P4_12_12$, respectively. These three phases are taken into account in the calculated profile.

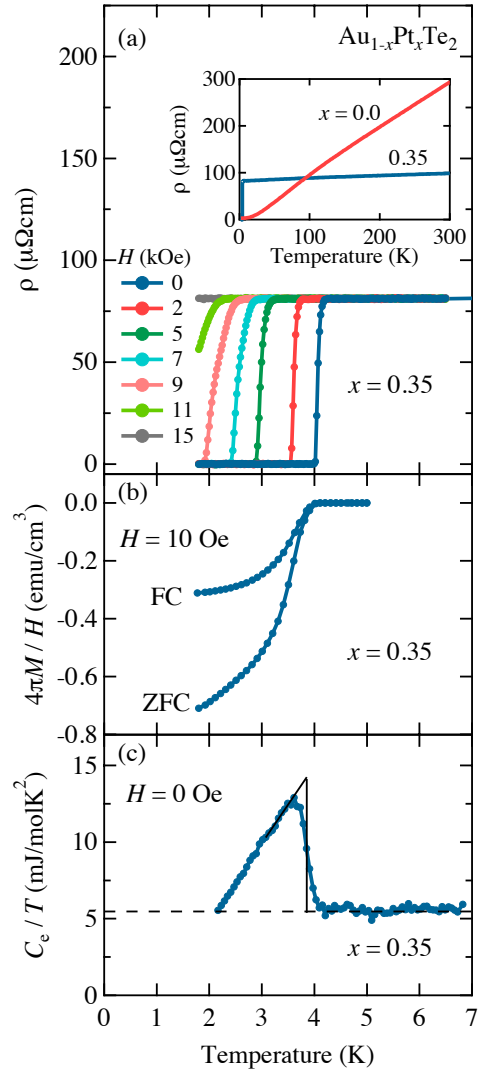

Fig. 3. (Color online) (a) Temperature dependence of the electrical resistivity ρ for $Au_{0.65}Pt_{0.35}Te_2$ at magnetic fields $H$ up to 15 kOe. The inset shows $\rho(T)$ for $AuTe_2$ and $Au_{0.65}Pt_{0.35}Te_2$ at a zero field in the temperature range of 1.8–300 K. (b) Temperature dependence of dc magnetization $M$ measured at $H = 10$ Oe for $Au_{0.65}Pt_{0.35}Te_2$ under conditions of zero-field cooling (ZFC) and field cooling (FC). (c) Temperature dependence of the electronic specific heat divided by the temperature, $C_e/T$, for $Au_{0.65}Pt_{0.35}Te_2$. $C_e$ was determined by subtracting the phonon contribution $\beta T^3 + \delta T^5$ from the total specific heat $C$, as shown in Fig. 5. The broken line corresponds to $\gamma$. The solid line represents an ideal jump at $T_c$, assuming entropy conservation at the transition.

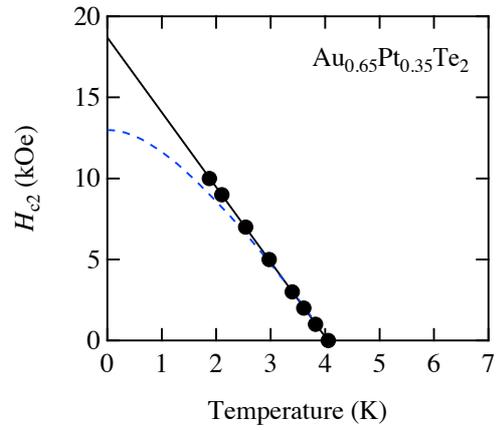

Fig. 4. (Color online) Temperature dependence of upper critical field $H_{c2}$ deduced from resistivity measurements. The solid line gives an estimate of the slope $-dH_{c2}/dT = 4.6$ kOe/K at $T = T_c$. The broken curve represents $H_{c2}(T)$ based on the Werthamer–Helfand–Hohenberg (WHH) theory[16].

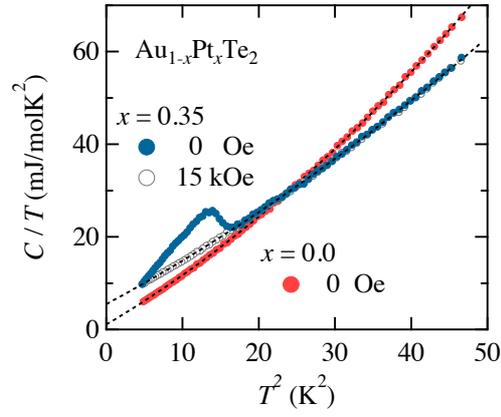

Fig. 5. (Color online) Specific heat divided by the temperature, $C/T$, as a function of $T^2$ for AuTe$_2$ at a zero field, and that for Au$_{0.65}$Pt$_{0.35}$Te$_2$ at a zero field and a magnetic field of 15 kOe. The broken curves denote the fits to $C/T = \gamma + \beta T^2 + \delta T^4$, where $\gamma$ is the electronic specific heat coefficient and $\beta$ and $\delta$ are those of phonon contributions.

**Supplementary information**

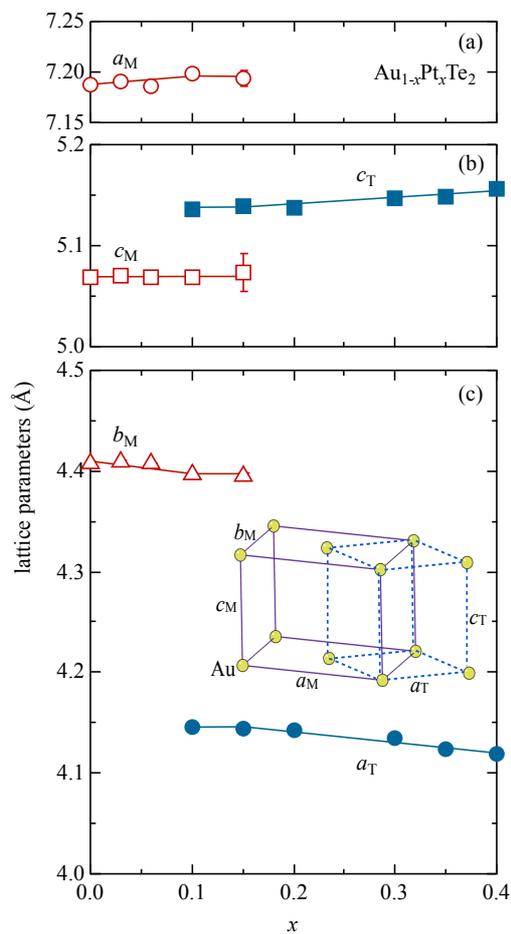

Supplementary Figure S1. $x$ dependences of lattice parameters for $Au_{1-x}Pt_xTe_2$ estimated from the powder X-ray diffraction pattern measured at room temperature. $a_M$, $b_M$, and $c_M$ indicate lattice parameters of the monoclinic structure (the average structure of $AuTe_2$) with the space group $C2/m$ (No. 12). $a_T$ and $c_T$ indicate lattice parameters of trigonal structure with the space group $P\bar{3}m1$ (No. 164). The inset shows the unit cell of the monoclinic (solid lines) and the trigonal (broken lines) structures.